\documentclass[english,aps,superscriptaddress,preprintnumbers,reprint,footinbib,amsmath,amssymb,prbr]{revtex4-2}
\usepackage[latin9]{inputenc}
\usepackage{color}
\usepackage{amsmath}
\usepackage{amssymb}
\usepackage{graphicx}
\usepackage{esint}

\usepackage{hyperref}
\hypersetup{
    colorlinks=true,
    linkcolor=blue,
    urlcolor=blue,
    citecolor=blue
        }
\makeatletter

\usepackage{color}

\@ifundefined{textcolor}{}{%
 \definecolor{BLACK}{gray}{0}
 \definecolor{WHITE}{gray}{1}
 \definecolor{RED}{rgb}{1,0,0}
 \definecolor{GREEN}{rgb}{0,1,0}
 \definecolor{BLUE}{rgb}{0,0,1}
 \definecolor{CYAN}{cmyk}{1,0,0,0}
 \definecolor{MAGENTA}{cmyk}{0,1,0,0}
 \definecolor{YELLOW}{cmyk}{0,0,1,0}
}

\usepackage{ulem}
\usepackage{aecompl}

\usepackage{epsfig}\usepackage{dcolumn}\usepackage{bm}
\usepackage{babel}

\makeatother

\usepackage{babel}
\begin{document}
\title{Topological charge Fano effect in multi-Weyl semimetals}
\author{W.C. Silva}
\affiliation{São Paulo State University (Unesp), School of Engineering, Department
of Physics and Chemistry, 15385-000, Ilha Solteira-SP, Brazil}
\author{W.N. Mizobata}
\affiliation{São Paulo State University (Unesp), School of Engineering, Department
of Physics and Chemistry, 15385-000, Ilha Solteira-SP, Brazil}
\author{J.E. Sanches}
\affiliation{São Paulo State University (Unesp), School of Engineering, Department
of Physics and Chemistry, 15385-000, Ilha Solteira-SP, Brazil}
\author{L.S. Ricco}
\affiliation{Science Institute, University of Iceland, Dunhagi-3, IS-107, Reykjavik,
Iceland}
\author{I.A. Shelykh}
\affiliation{Science Institute, University of Iceland, Dunhagi-3, IS-107, Reykjavik,
Iceland}
\affiliation{ITMO University, St. Petersburg, 197101, Russia}
\author{M. de Souza}
\affiliation{São Paulo State University (Unesp), IGCE, Department of Physics, 13506-970,
Rio Claro-SP, Brazil}
\author{M.S. Figueira}
\affiliation{Instituto de Física, Universidade Federal Fluminense, 24210-340, Niterói,
Rio de Janeiro, Brazil}
\author{E. Vernek}
\affiliation{Instituto de Física, Universidade Federal de Uberlândia, Uberlândia,
38400-902, Minas Gerais, Brazil}
\author{A.C. Seridonio}
\email[corresponding author: ]{antonio.seridonio@unesp.br}

\affiliation{São Paulo State University (Unesp), School of Engineering, Department
of Physics and Chemistry, 15385-000, Ilha Solteira-SP, Brazil}
\begin{abstract}
We theoretically analyze the Fano interference in a single impurity
multi-Weyl semimetal hybrid system and show the emergence of the topological
charge Fano effect in the bulk local density of states. In multi-Weyl
semimetals, the number of Fermi arcs at the system boundaries is determined
by the topological charge $J$, a direct consequence of the ``bulk-boundary''
correspondence principle. Analogously, we find that $J$ also modulates
the bulk Fano profile of the system with an embedded quantum impurity.
Thus, by increasing $J$, the Fano lineshape evolves from resonant,
typical for $J=1$ (single Weyl), towards antiresonant, extrapolating
to the so-called hyper Weyl semimetals with $J\gg1$. Specially for
the maximum case protected by the rotational symmetry $C_{2J=6}$,
namely the $J=3$ (triple Weyl), which acquires asymmetric Fano profile,
the Fano parameter absolute value is predicted to be $\tan(C_{2J=6})$,
where $C_{2J}\equiv(360^{\circ}/2J)$ defines the rotational angle.
Hence, the Fano discretization in the $J$ term introduces the topological
charge Fano effect in multi-Weyl semimetals. {We also
suggest a transport device where we expect that the proposed Fano
effect could be detected.}
\end{abstract}
\maketitle

\section{Introduction}

Multi-Weyl semimetals{\cite{Hasan2021,Hasan2017,Yan,Zheng,Armitage,Hu,Marques,MizobataWeyl}
are intriguing generalizations of standard Weyl semimetals\cite{Park,Dantas,Hayata,Xu,Liu,Pedrosa,Fang,Dantas2018,Ahn,Mukherjee,Lu2019},
once they can lead to a plethora of fascinating effects, such as chiral,
optical and transport anomalous properties\cite{Hayata,Ahn,Mukherjee,Chen,Marques,MizobataWeyl,Park,Dantas2018,Nag_2020,https://doi.org/10.48550/arxiv.2203.12756}.
In multi-Weyl semimetals, the band-structures at the so-called Weyl
crossing points, show highly anisotropic dispersion relations, being
relativistic exclusively in one momentum direction, while in the other
two, a power-law dependence is ruled by the topological charge $J$\cite{Lu2019}.
This topological number corresponds to the quantized Berry phase of
the Dirac fermions in graphene\cite{Neto2009}. As Weyl points appear
in pairs with opposite chiralities, they behave as source and drain
of an Abelian Berry curvature, thus mimicking (anti)monopoles placed
far apart in the reciprocal space. Amazingly, such points are connected
to each other via crystal boundaries, in particular, by opened surface
states known as Fermi arcs. Notably, these exotic states can be observed
by ARPES\cite{Hasan2021,Hasan2017} and turn into the experimental
proof of the magnetic monopoles existence in the momenta space.

In multi-Weyl semimetals, the winding number also plays the role of
a higher topological charge $J>1$\cite{Hasan2021}. This charge arises
from the merging of $J$ single chiral-degenerate Weyl nodes into
multi-Weyl points, which are point group symmetry protected up to
$J=3$\cite{Fang}, namely, by means of the rotational symmetry $C_{2J}$.
In this manner, the ``bulk-boundary'' correspondence dictates that
for a given $J$ value determined in an infinite bulk system, $J$
pairs of Fermi arcs appear in the corresponding finite version of
the setup\cite{Dantas}. Some examples of multi-Weyl materials are
$\text{{HgCr}}_{2}\text{{Se}}_{4}$ and $\text{{SrSi}}_{2}$ with
$J=2$ (double Weyl)\cite{Fang,Xu,Huang,Chen}, and $\text{{A(MoX)}}_{3}$
(A=Rb or Tl and X=Te) with $J=3$ (triple Weyl)\cite{Liu}.

In this work, we focus on the ``bulk-boundary'' correspondence for
Fermi arcs surface states and the topological charge from the bulk,
in order to present the concept of the topological charge Fano effect
in multi-Weyl semimetals. To this end, we theoretically explore the
bulk Fano interference\cite{Fano1961,Miroshnickenko} in the LDOS
(\textit{local density of states}) for a single impurity multi-Weyl
semimetal hybrid system, as sketched in Fig.\ref{fig:Pic1}(a). As
a matter of fact, the Fano interference arises from the coupling between
a discrete energy level and an energy continuum\cite{Miroshnickenko}.
It has been widely investigated in several platforms, ranging from
classical mechanics\cite{Joe_2006} to topological superconductivity\cite{Xia,Ricco}.
Coupled harmonic oscillators with a driving force\cite{Joe_2006},
photonic systems\cite{Limonov}, Jaynes-Cummings-like cavities\cite{Jakob},
electronic quantum transport setups made of Anderson adatoms\cite{Anderson,Madhavan,Knorr,FanoWithAdatoms3},
atomically frustrated molecules in Weyl metals\cite{MizobataWeyl}
and topological superconducting nanowires with quantum dots\cite{Ricco},
among others\cite{Miroshnickenko}, constitute the broad variety of
examples where Fano interference manifests itself.

Here, we reveal that the increase of $J$ modifies the bulk Fano profile
of a multi-Weyl semimetal with a single impurity, by means of the
tuning of Fano asymmetry parameter $q_{J}$, from resonant lineshape
$(\left|q_{J=1}\right|\rightarrow\infty)$ towards antiresonant one
$(\left|q_{J\gg1}\right|\rightarrow0)$. We highlight that while the
former identifies $J=1$ case (single Weyl) {[}inset panel of Fig.\ref{fig:Pic1}(b){]},
the latter predicts a Fano antiresonant profile characterized by $J\gg1$.
For such a case, we relax, as we shall clarify later on, the aforementioned
crystalline protection\cite{Fang} and make explicit that the fingerprint
for this situation, which we introduce as the hyper Weyl semimetal,
is represented by a suppressed Fano parameter $(\left|q_{J\gg1}\right|\rightarrow0)$
{[}inset panel of Fig.\ref{fig:Pic1}(d){]}.

We clarify that hyper Weyl semimetals should be understood as a conjecture,
being a hypothetical case corresponding to a huge topological charge.
However, some research groups have reported spinless platforms with
$J=4$\cite{1J4,2J4}, pointing out that it is still capital to consider
a generalized description. Noteworthy, for $J\geq3,$ the Fano parameter
becomes finite, discretized in $J$ and shows a decaying behavior.
Particularly for the maximum allowed case by the point symmetry group
protection, i.e., the $J=3$ value for the $C_{2J=6}$ rotational
symmetry group, we predict $\left|q_{J=3}\right|=\tan(C_{2J=6})$,
with the rotational angle $C_{2J}\equiv(360^{\circ}/2J)$ and an asymmetric
Fano lineshape {[}inset panel of Fig.\ref{fig:Pic1}(c){]}. Thereby,
our findings introduce the idea of the topological charge Fano effect
in multi-Weyl semimetals.

\section{The Model}

The Hamiltonian mimicking our system {[}Fig.\ref{fig:Pic1}(a){]},
with $\hbar=1$, reads

\begin{equation}
\mathcal{H}=\mathcal{H}_{\text{{Weyl}}}+\mathcal{H}_{\text{{Imp.}}}+\mathcal{H}_{\text{{Hyb.}}},\label{eq:TotalH}
\end{equation}
where

\begin{equation}
\mathcal{H}_{\text{{Weyl}}}=\sum_{\textbf{k}s}\psi_{\textbf{k}s}^{\dagger}s[D(\tilde{k}_{-}^{J}\sigma_{+}+\tilde{k}_{+}^{J}\sigma_{-})+v_{F}(k_{z}-sQ)\sigma_{z}]\psi_{\textbf{k}s}\label{eq:Weyl}
\end{equation}
is the part describing multi-Weyl fermions with spinor $\psi_{\textbf{k}s}^{\dagger}=(c_{\textbf{k}s\uparrow}^{\dagger}c_{\textbf{k}s\downarrow}^{\dagger})$,
$c_{\textbf{k}s\sigma}^{\dagger}$ ($c_{\textbf{k}s\sigma}$) for
creation (annihilation) of an electron carrying winding number $J$,
momentum $\textbf{k}$, spin $\sigma=\uparrow,\downarrow$ and chirality
$s=\pm1$ for the Weyl nodes $sQ$, which break time-reversal symmetry.
The nonrelativistic part is expressed in terms of $\tilde{k}_{\pm}=(k_{x}\pm ik_{y})/k_{D}$,
being $k_{D}=D/v_{F}$ $(D)$ the \textit{Debye-like} momentum (energy)
cutoff as in graphene system\cite{Uchoa}, $\sigma_{\pm}=\frac{1}{2}(\sigma_{x}\pm i\sigma_{y})$
and $\sigma_{z}$ are the Pauli matrices. We stress that the winding
number $J$, namely the topological charge, gives the number of Fermi
arcs pairs at the system boundaries, as ensured by the ``bulk-boundary''
correspondence principle\cite{Dantas}. In the last term of Eq.(\ref{eq:Weyl}),
which is of relativistic-type, the slope of the Dirac cones in the
$z$-direction of the momentum space is the Fermi velocity $v_{F}$.

\begin{figure}[!]
\centering\includegraphics[width=1\columnwidth]{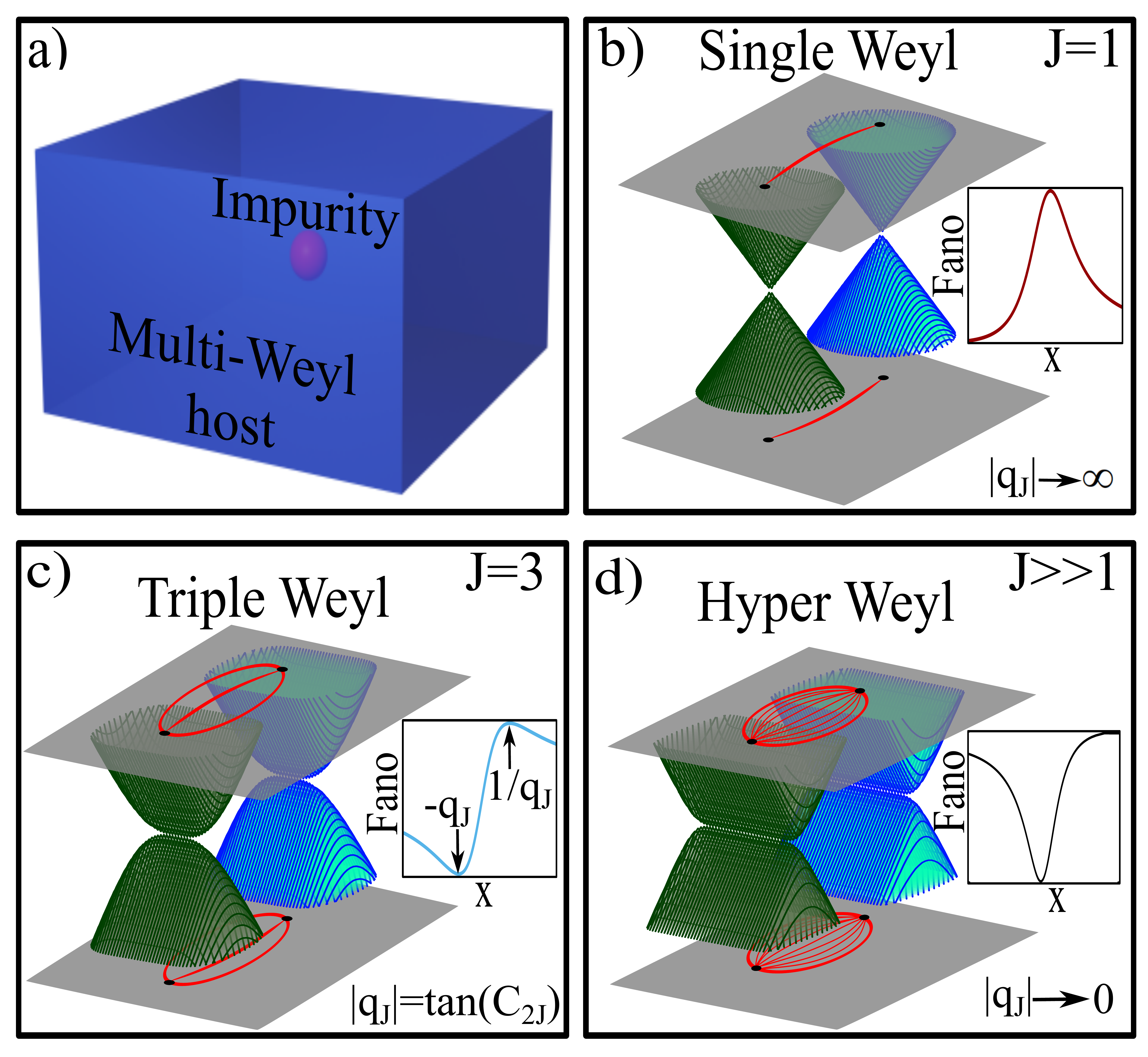}\caption{\label{fig:Pic1} {(Color online) Overview of the topological charge
Fano effect in multi-Weyl semimetals: (a) Slab of a bulk multi-Weyl
semimetal system hosting an impurity. (b)-(d) Qualitative summary
of our findings: corresponding band-structures from Eq.(\ref{eq:DispersionRelation})
with $k_{y}=0$, $Q=0.4k_{D}$ and Fermi arcs surface states upon
changing the topological charge $J$. Related bulk Fano profiles versus
the dimensionless resonant energy detuning $x$ of the impurity appear
depicted at the inset panels. The Fano profile evolves from resonant
behavior {[}panel (b){]} to the antiresonant-type {[}panel (d){]}
as $J$ increases. The pairs of Fermi arcs are determined by the $J$
value, which imposes the bulk Fano lineshape. This profile is determined
by the absolute value of the Fano asymmetry parameter $\left|q_{J}\right|=\tan(C_{2J})$
{[}panel (c){]}, where $C_{2J}\equiv(360^{\circ}/2J)$ stands for
the angle of the rotational symmetry group. In summary, the ``bulk-boundary''
correspondence\cite{Dantas} defines the grounds of the topological
charge Fano effect in multi-Weyl systems.}}
\end{figure}

The band-structure of Eq.(\ref{eq:Weyl}) can be computed straightforwardly
and leads to the following dispersion relation
\begin{equation}
\varepsilon_{\textbf{k}s}^{\pm}=\pm v_{F}\sqrt{k_{zs}^{2}+|\tilde{k}_{+}|^{2J}k_{D}^{2}},\label{eq:DispersionRelation}
\end{equation}
wherein $+(-)$ corresponds to the conduction (valence) band, with
$k_{zs}=k_{z}-sQ$ and it is depicted in Figs.\ref{fig:Pic1}(b)-(d).
Additionally, it is worth mentioning that for $J=1$ (single Weyl)
the Weyl semimetal has well-defined Dirac cones in all momentum directions
{[}Fig.\ref{fig:Pic1}(b){]}, while for $J\gg1$ we have a hypothetical
hyper Weyl semimetal case {[}Fig.\ref{fig:Pic1}(d){]}, in which its
band-structure shape saturates due to a huge topological charge. Further,
a single Anderson-like impurity\cite{Anderson} can be described by
the Hamiltonian
\begin{align}
\mathcal{H}_{\text{{Imp.}}} & =-\frac{U}{2}+\sum_{\sigma}(\varepsilon_{d\sigma}+\frac{U}{2})n_{d\sigma}+\frac{U}{2}(\sum_{\sigma}n_{d\sigma}-1)^{2},\nonumber \\
\label{eq:Imp}
\end{align}
where the impurity electronic energy level is $\varepsilon_{d\sigma},$
with number operator $n_{d\sigma}=(n_{d\sigma})^{2}=d_{\sigma}^{\dagger}d_{\sigma}$,
being $d_{\sigma}^{\dagger}$ ($d_{\sigma}$) the corresponding creation
(annihilation) operator and $U$ is the Coulomb repulsion between
two electrons with opposite spins $(\bar{\sigma}=-\sigma)$. The hybridization
term, which accounts for the host-impurity coupling, reads

\begin{equation}
\mathcal{H}_{\text{{Hyb.}}}=v\sum_{\sigma}(f_{0\sigma}^{\dagger}d_{\sigma}+\text{{H.c.}),}\label{eq:Hyb}
\end{equation}
where the field operator

\begin{equation}
f_{0\sigma}=\frac{1}{\sqrt{\mathcal{N}}}\sum_{\textbf{k}s}c_{\textbf{k}s\sigma}\label{eq:fo}
\end{equation}
describes the host site locally coupled to an embedded quantum impurity,
with $v$ being the impurity-host coupling strength and $\mathcal{N}$
the number of states delimited by $k_{D}$.

We would like to call attention to the following: by making the choice
$\varepsilon_{d\sigma}=-\frac{U}{2}$, the second term of Eq.(\ref{eq:Imp})
disappears and the Hamiltonian $\mathcal{H}$ becomes invariant under
the particle-hole transformation $c_{\textbf{k}s\sigma}\rightarrow c_{-\textbf{k}s\sigma}^{\dagger}$
and $d_{\sigma}\rightarrow-d_{\sigma}^{\dagger}$. This characterizes
the particle-hole symmetric regime of the model, which will be employed
without loss of generality, in order to determine the system bulk
LDOS. Consequently, the LDOS profile exhibits mirror symmetry in the
energy domain $\varepsilon$ and the bulk Fano profile can be finally
known.

\section{LDOS and Fano profile}

From the time Fourier transform of the retarded Green's function (GF)

\begin{equation}
\mathcal{G}_{\sigma}=-i\theta(t)\left\langle \left\{ f_{0\sigma}(t),f_{0\sigma}^{\dagger}(0)\right\} \right\rangle _{\mathcal{H}},\label{eq:GFfo}
\end{equation}
 i.e. $\tilde{\mathcal{G}}_{\sigma}$, we verify the validity of the
Dyson equation

\begin{equation}
\tilde{\mathcal{G}}_{\sigma}=\tilde{\mathcal{G}}_{\sigma}^{0}+\tilde{\mathcal{G}}_{\sigma}^{0}v\tilde{\mathcal{G}}_{\text{{Imp.}}\sigma}v\tilde{\mathcal{G}}_{\sigma}^{0}\label{eq:GFtilfo}
\end{equation}
via the equation-of-motion approach\cite{Flensberg}, with $\mathcal{G}_{\sigma}^{0}$
and

\begin{equation}
\mathcal{G}_{\text{{Imp.}}\sigma}=-i\theta(t)\left\langle \left\{ d_{\sigma}(t),d_{\sigma}^{\dagger}(0)\right\} \right\rangle _{\mathcal{H}}\label{eq:GFimp}
\end{equation}
representing the pristine multi-Weyl and impurity GFs, respectively.
Thus, the Fano formula\cite{Fano1961,Miroshnickenko} in the bulk

\begin{equation}
\text{{LDOS}}=-\frac{1}{\pi}\text{{Im}}\tilde{\mathcal{G}}_{\sigma}\label{eq:LDOS}
\end{equation}
is expected to emerge, if in the impurity

\begin{equation}
\text{{DOS}}=-\frac{1}{\pi}\text{{Im}}\tilde{\mathcal{G}}_{\text{{Imp.}}\sigma},\label{DOS}
\end{equation}
their resonant states\cite{Flensberg} exhibit a lorentzian profile.

As we consider the case of $T\ll T_{\text{{K}}}\rightarrow0$ (Kondo
temperature)\cite{Hewson} and the system has a pseudogap at the Fermi
level, Kondo correlations do not emerge\cite{Pedrosa} and consequently,
the Coulomb blockade regime\cite{Flensberg} takes place. The latter
is characterized solely by the resonant states $\varepsilon_{d\sigma}$
and $\varepsilon_{d\sigma}+U$, and the correlation $U$ in $\tilde{\mathcal{G}}_{\text{{Imp.}}\sigma}$,
can be safely treated in the framework of the Hubbard-I approximation\cite{Flensberg,Marques,MizobataWeyl}.
{The Hubbard-I approximation is indeed, a mean-field
calculation, i.e., a truncation scheme on the system GFs, which determines
the impurity GF $\tilde{\mathcal{G}}_{\text{{Imp.}}\sigma},$ in particular,
by accounting for the electronic correlation $U$ in Eq.(\ref{eq:Imp})
within a certain regime of validity. We stress that the presence of
the Hubbard term $U$ in Eq.(\ref{eq:Imp}), which shows a quadratic
dependence on the number operator $n_{d\sigma},$ prevents inevitably,
the analytical and exact evaluation of $\tilde{\mathcal{G}}_{\text{{Imp.}}\sigma}.$
This lack of completeness, naturally, does not catch the complete
low-energy regime of the single impurity Anderson model\cite{Anderson}.
More specifically, the one characterized by $T\ll T_{\text{{K}}},$
$\varepsilon_{d\sigma}<0,$ $\varepsilon_{d\sigma}+U>0$ and, as a
result, the Kondo peak present in Eq.(\ref{DOS}). We call the attention
that such a resonance is a many-body effect, which is due to a spin-flip
process between the electrons from the impurity and the host conduction
states. }

{It is worth mentioning that one of us in Ref.\cite{Pedrosa}
has demonstrated, by employing the Numerical Renormalization Group\cite{Pedrosa},
that the Kondo peak emerges solely in multi-Weyl semimetals when the
Fermi level is off resonance from the Dirac point, i.e., the so-called
charge neutrality point $\varepsilon=0$. By approaching this spot,
the multi-Weyl semimetal presents a pseudogap, as we will verify later
on, that scales with the power-law $(\varepsilon^{2})^{1/J}$ in the
topological charge $J$ for the pristine host density of states. It
means that at the charge neutrality point, the host does not contain
states to screen in an antiferromagnetic way the localized magnetic
moment at the impurity site and lead to the Kondo peak in the impurity
density of states of Eq.(\ref{DOS}). Thus, the spin-flip process
quenches and even with $T\ll T_{\text{{K}}}\rightarrow0,$ for multi-Weyl
semimetals, the Kondo peak does not rise at $\varepsilon=0.$ This
scenario is fully distinct from a metallic system, once at the corresponding
charge neutrality point, the pristine host density of states is finite.
As the Hubbard-I method disregards such a spin-flip mechanism to obtain
the impurity GF $\tilde{\mathcal{G}}_{\text{{Imp.}}\sigma},$ then
we can safely adopt it to our system, only if we maintain the Fermi
level at $\varepsilon=0,$ where for multi-Weyl semimetals the absence
of states is ensured. By taking into account such an assumption, we
employ the well-established GF in the Hubbard-I approximation as follows:
\begin{equation}
\tilde{\mathcal{G}}_{\text{{Imp.}}\sigma}=-\frac{1}{v^{2}\text{{Im}\ensuremath{\tilde{\mathcal{G}}_{\sigma}^{0}}}}\left(\frac{w_{x}}{x+i}+\frac{w_{\bar{x}}}{\bar{x}+i}\right).\label{eq:Hubbard}
\end{equation}
}This ensures, as expected, the lorentzian lineshape in the $\text{{DOS}},$
with $w_{x}=1-\left\langle n_{d\bar{\sigma}}\right\rangle $ and $w_{\bar{x}}=1-w_{x}$
being spectral weights for the dimensionless resonant energies detuning

\begin{equation}
x=\frac{\varepsilon-\varepsilon_{d\sigma}-v^{2}\text{Re}\tilde{\mathcal{G}}_{\sigma}^{0}}{-v^{2}\text{{Im}\ensuremath{\tilde{\mathcal{G}}_{\sigma}^{0}}}}\label{eq:naturalx}
\end{equation}
and

\begin{equation}
\bar{x}=\frac{\varepsilon-\varepsilon_{d\sigma}-U-v^{2}\text{Re}\tilde{\mathcal{G}}_{\sigma}^{0}}{-v^{2}\text{{Im}\ensuremath{\tilde{\mathcal{G}}_{\sigma}^{0}}}},\label{eq:naturalxbar}
\end{equation}
respectively, wherein

\begin{equation}
\left\langle n_{d\sigma}\right\rangle =\int_{-D}^{0}\text{{DOS}}d\varepsilon\label{eq:number}
\end{equation}
is the impurity occupation. {The pristine host GF,
or simply the propagator of the pristine Weyl fermions, is just}

{
\begin{align}
\tilde{\mathcal{G}}_{\sigma}^{0} & =\frac{1}{\mathcal{N}}\sum_{\textbf{k}s}\frac{\varepsilon+i0^{+}}{(\varepsilon+i0^{+})^{2}+(\varepsilon_{\textbf{k}s}^{+})^{2}}=\text{{Re}\ensuremath{\tilde{\mathcal{G}}_{\sigma}^{0}}}+i\text{{Im}\ensuremath{\tilde{\mathcal{G}}_{\sigma}^{0}}}\nonumber \\
 & =\left|\tilde{\mathcal{G}}_{\sigma}^{0}\right|\exp(i\delta_{J}),\label{eq:PHostGF}
\end{align}
where $\delta_{J}$ represents the phase of the propagator, in particular
in the absence of the impurity, which is expected to depend upon the
topological charge $J$ via the dispersion relation $\varepsilon_{\textbf{k}s}^{+}$
of Eq.(\ref{eq:DispersionRelation}). It reads
\begin{equation}
\tan\delta_{J}=\frac{\text{{Im}\ensuremath{\tilde{\mathcal{G}}_{\sigma}^{0}}}}{\text{{Re}\ensuremath{\tilde{\mathcal{G}}_{\sigma}^{0}}}}.\label{eq:TanDelta}
\end{equation}
This quantity, as we will see later on, is deeply connected to the
Fano asymmetry parameter $q_{J}$ of the system, responsible for modulating
the bulk Fano profile for the LDOS. Particularly for a multi-Weyl
semimetal with $J\geq3$, in addition, we will show that $q_{J}$
becomes ruled by the angle $(360^{\circ}/2J),$ which surprisingly,
is recognized as the angle of the rotational symmetry group $C_{2J}.$
As we know, such a symmetry group stabilizes locally multi-Weyl points
in the momentum space\cite{Fang}. }

Now we are able to express the LDOS according to Fano formula\cite{Fano1961,Miroshnickenko}.
We begin by introducing into {Eq.(\ref{eq:GFtilfo})},
the quantities as follows:

\begin{align}
\text{{Re}\ensuremath{\tilde{\mathcal{G}}_{\sigma}^{0}}} & =-\frac{1}{\pi}\int_{-D}^{+D}\frac{\text{{Im}}\tilde{\mathcal{G}}_{\sigma}^{0}}{\varepsilon-y}dy\nonumber \\
 & =-\frac{1}{\pi}\text{{sgn}}(\varepsilon)\text{{Im}}\tilde{\mathcal{G}}_{\sigma}^{0}\int_{-D/\varepsilon}^{+D/\varepsilon}\frac{(u^{2})^{1/J}}{1-u}du\nonumber \\
 & =-q_{J}\text{{Im}}\tilde{\mathcal{G}}_{\sigma}^{0},\label{eq:KramersKronig}
\end{align}
due to the Kramers-Kronig relations\cite{Flensberg}, wherein $y=u\varepsilon$
and the Fano asymmetry parameter is given by{
\begin{align}
q_{J} & =-\frac{\text{{Re}\ensuremath{\tilde{\mathcal{G}}_{\sigma}^{0}}}}{\text{{Im}}\tilde{\mathcal{G}}_{\sigma}^{0}}=-\cot\delta_{J},\nonumber \\
 & =\frac{1}{\pi}\text{{sgn}}(\varepsilon)\text{{P.V.}}\int_{-D/\varepsilon}^{+D/\varepsilon}\frac{(u^{2})^{1/J}}{1-u}du,\label{eq:FanoqGeral}
\end{align}
where \text{P.V.} stands for the Cauchy principal value and} {we
clearly see that the phase $\delta_{J}$ of Eq.(\ref{eq:TanDelta})
for the pristine Weyl fermions propagator of Eq.(\ref{eq:PHostGF})
then dictates the Fano asymmetry parameter. Moreover,}
\begin{equation}
\text{{Im}}\tilde{\mathcal{G}}_{\sigma}^{0}=-\frac{3\pi^{3/2}\Gamma(\frac{1}{J})}{2JD^{\frac{J+2}{J}}\Gamma(\frac{2+J}{2J})}(\varepsilon^{2})^{1/J},\label{eq:ImG0}
\end{equation}
with $\Gamma(x)$ being the Gamma function and the power-law $(\varepsilon^{2})^{1/J}$
is characterized by a pseudogap at the Fermi level ($\varepsilon=0$).

We emphasize that Eq.(\ref{eq:ImG0}) holds for arbitrary $J$ and
mention that so far, solely analytical expressions up to $J=3$ were
obtained\cite{Lu2019}. We are aware that the crystalline rotational
symmetry $C_{2J}$ imposes the limitation $J\leqslant3$, in particular
when the spin degree of freedom comes into play\cite{Fang}. However,
the $J=4$ case is still possible and emerges in spinless systems\cite{1J4,2J4}.
Thus, we develop an extrapolation given by Eq.(\ref{eq:ImG0}) and
get a generalized Fano asymmetry parameter.

The aforementioned accomplishment was possible after employing in
{Eq.(\ref{eq:PHostGF})} the procedures as follows:
(i) the standard substitution $\mathcal{N}=\sum_{\textbf{k}s}\rightarrow\frac{\varOmega}{(2\pi)^{3}}\int d^{3}\textbf{k}=\frac{\varOmega}{6\pi^{2}}k_{D}^{3}$,
with $\Omega$ as the volume element in real space; (ii) the hyper-spherical
transformation given by $k_{x}=k_{D}(\frac{\varepsilon_{\textbf{k}s}^{+}\sin\theta}{D})^{\frac{1}{J}}\cos\phi$,
$k_{y}=k_{D}(\frac{\varepsilon_{\textbf{k}s}^{+}\sin\theta}{D})^{\frac{1}{J}}\sin\phi$
and $k_{zs}=k_{D}\frac{\varepsilon_{\textbf{k}s}^{+}}{D}\cos\theta$
($0\leq\theta\leq\pi,$ $0\leq\phi\leq2\pi$), with Jacobian

\begin{equation}
\text{{J}}(\varepsilon_{\textbf{k}s}^{+},\theta,\phi)=\frac{k_{D}^{3}}{D}(\frac{\varepsilon_{\textbf{k}s}^{+}}{D})^{2/J}\frac{(\sin\theta)^{\frac{2}{J}-1}}{J}\label{eq:Jacobian}
\end{equation}
and property $\int\tilde{\mathcal{G}}_{\sigma}^{0}d^{3}\textbf{k}=\int\tilde{\mathcal{G}}_{\sigma}^{0}\text{{J}}(\varepsilon_{\textbf{k}s}^{+},\theta,\phi)d\varepsilon_{\textbf{k}s}^{+}d\theta d\phi$.

We can finally obtain the bulk Fano profile, which from here, we call
by natural Fano profile (NFP), once it is expressed in terms of their
natural coordinates $x$ and $\bar{x}$. Taking into account the spin
degree of freedom, we find the $\text{{NFP}=}2\text{{LDOS}}/\rho_{0}(1+q_{J}^{2})$,
with

\begin{equation}
\rho_{0}=-\frac{1}{\pi}\text{{Im}}\tilde{\mathcal{G}}_{\sigma}^{0}\label{eq:DOShost}
\end{equation}
as the pristine multi-Weyl DOS and
\begin{equation}
\text{{NFP}}=\frac{2}{1+q_{J}^{2}}\left[w_{x}\frac{(x+q_{J})^{2}}{x^{2}+1}+w_{\bar{x}}\frac{(\bar{x}+q_{J})^{2}}{\bar{x}^{2}+1}\right],\label{eq:Normalized Fano Profile}
\end{equation}
which holds in the wide-band limit $D/\varepsilon\rightarrow\infty$.
As we are interested in impurity levels nearby the Fermi energy, such
a limit prevents that the time-reversal symmetry breaking lifts the
system spin degeneracy\cite{Pedrosa}. From Eq. (\ref{eq:Normalized Fano Profile})
and for $\varepsilon<0$ ($\varepsilon>0$), the NFP shows amplitudes
of minimum and maximum at $x=-q_{J}$ ($\bar{x}=-q_{J}$) and $x=1/q_{J}$
($\bar{x}=1/q_{J}$), respectively.

We highlight that the LDOS spectral lineshape itself, as we shall
see in the numerical analysis, will not exhibit a Fano profile as
a function of energy $\varepsilon$ for a given $q_{J}$, as it occurs
for flat band systems with energy and $J$ independent host DOS\cite{Madhavan,Knorr,FanoWithAdatoms3}.
Additionally, we will verify that for the revealing of such a behavior,
one should analyze the Fano profile as a function of $x$ or $\bar{x}$,
namely, the natural coordinates for the Fano profile to emerge. Nevertheless,
before that, we should firstly evaluate carefully the integral over
$u$ variable in Eq.(\ref{eq:FanoqGeral}).

We call attention that, in particular for $q_{J=1}$ and $q_{J=2}$,
the functions depending on $u$ do not vanish in the limits $u\rightarrow\pm\infty$,
which is a common pathology in low-energy models\cite{Uchoa}. This
feature constitutes a technical difficulty in solving Eq.(\ref{eq:FanoqGeral})
numerically. Hence, to handle accordingly with this lack of integrability
issue, we should, as already performed in graphene system\cite{Uchoa},
solve first the integral analytically by keeping the ratio $D/\varepsilon$
finite and assuming later on, the limit $\varepsilon/D\ll1$ in the
$\text{{Re}\ensuremath{\tilde{\mathcal{G}}_{\sigma}^{0}}}$ evaluations.
Such cases are then described by
\begin{equation}
\text{{Re}\ensuremath{\tilde{\mathcal{G}}_{\sigma}^{0}}}(J=1)=\frac{3\varepsilon}{D^{3}}(\varepsilon\ln\frac{|D+\varepsilon|}{|D-\varepsilon|}-2D)\label{eq:J1}
\end{equation}
and
\begin{equation}
\text{{Re}\ensuremath{\tilde{\mathcal{G}}_{\sigma}^{0}}}(J=2)=\frac{3\pi}{4D^{2}}\varepsilon\ln\frac{\varepsilon^{2}}{|\varepsilon^{2}-D^{2}|},\label{eq:J2}
\end{equation}
respectively, which provide

{
\begin{equation}
q_{J\leq2}=-\frac{\text{{Re}\ensuremath{\tilde{\mathcal{G}}_{\sigma}^{0}}}(J\leq2)}{\text{{Im}}\tilde{\mathcal{G}}_{\sigma}^{0}(J\leq2)}=-\cot\delta_{J\leq2}\label{eq:FanoUpTo2}
\end{equation}
}as dependent both on energy for $\varepsilon/D\ll1$ and topological
charge $J$.

However, the necessary vanishing behavior is present in $q_{J\geq3}$
as a function of $u$ and consequently, it makes the integral in Eq.(\ref{eq:FanoqGeral})
to behave finite for $D/\varepsilon\rightarrow\infty$, which can
be found analytically simultaneously with the ratio $D/\varepsilon\rightarrow\infty$.
This results into an interesting finding, i.e., an energy-independent
Fano asymmetry parameter discretized in the topological charge $J$,
which reads
\begin{equation}
q_{J\geq3}=-\text{{sgn}}(\varepsilon)\tan(C_{2J\geq6}),\label{eq:FanoqMaior3}
\end{equation}
where we define $C_{2J}\equiv(360^{\circ}/2J)$ as the angle for the
corresponding rotational symmetry group. Together with Eqs.(\ref{eq:J1}),
(\ref{eq:J2}) and (\ref{eq:FanoUpTo2}), this gives the set of analytical
expressions that defines the topological charge Fano effect in multi-Weyl
systems. Notice that for $J=3$, $q_{J=3}=-\text{{sgn}}(\varepsilon)\tan(C_{2J=6})=-\text{{sgn}}(\varepsilon)\sqrt{3}$,
while for $J\gg1$ we have $\left|q_{J\gg1}\right|\rightarrow0,$
which corresponds to the maximum allowed point group symmetry protected
case, namely, the $C_{2J=6}$ rotational symmetry group, and the hypothetical
hyper Weyl semimetal, respectively.

{We emphasize that for $J\leq3$ such a crystalline
symmetry, hereby expressed in Eq.(\ref{eq:FanoqMaior3}) via the $C_{2J}$
parameter, then stabilizes the merge of chiral-degenerate Weyl points
with $J=1$ each and leads to an unique point enclosing a multi-topological
charge $J>1.$ This spot in momentum space is the so-called multi-Weyl
point, where the aforementioned symmetry glues together multiple Weyl
points with unitary topological charge and prevent them to split away. }

{Thus, we can point out that the Fano parameter role
in the accounted picture is self-contained in Eq.(\ref{eq:FanoqMaior3})
and arises from Eq.(\ref{eq:TanDelta}), which authorizes the link
$q_{J\geq3}=-\cot\delta_{J\geq3}=-\text{{sgn}}(\varepsilon)\tan(C_{2J\geq6}),$
from where we perceive that the NFP of Eq.(\ref{eq:Normalized Fano Profile})
becomes settled by the topological charge $J.$ As for $J\geq3$ the
phase $\delta_{J\geq3}$ of the Weyl fermions in the propagator of
Eq.(\ref{eq:PHostGF}) depends on the dispersion relation $\varepsilon_{\textbf{k}s}^{+}$
of Eq.(\ref{eq:DispersionRelation}), it yields the Fano asymmetry
parameter $q_{J\geq3}$ to be ruled by the angle $(360^{\circ}/2J),$
which is related to the rotational symmetry group $C_{2J\geq6}.$
Distinctly, for $J\leq2$ the dispersion relation $\varepsilon_{\textbf{k}s}^{+}$
introduces in $q_{J\leq2}$ of Eq.(\ref{eq:FanoUpTo2}) complex dependencies
on the topological charge $J,$ in particular obeying Eqs.(\ref{eq:J1})
and (\ref{eq:J2}), which affect peculiarly the NFP of Eq.(\ref{eq:Normalized Fano Profile}). }

{As aftermath, independently of the $J$ strength,
the broadening of the impurity levels {[}Eq.(\ref{eq:Hubbard}){]}
and the LDOS change into a Fano-type profile in the natural coordinates
{[}Eq.(\ref{eq:Normalized Fano Profile}){]}, are expected to occur
in both the scenarios. In summary, the system exhibits two paths of
transport that interfere to each other: one consists of electrons
that travel through the orbital $f_{0\sigma}$ of the host and that
wherein they visit the impurity $d_{\sigma}$ and return to $f_{0\sigma},$
being a process modulated by $q_{J}.$}

\section{Results and discussion}

\subsection{Natural Fano profile and discretized Fano parameter}

As stated previously, we consider the particle-hole symmetric regime.
In this case, $\varepsilon_{d\sigma}=-\frac{U}{2}$ and $w_{x}=w_{\bar{x}}=1/2$
($\left\langle n_{d\bar{\sigma}}\right\rangle =1/2$ from self-consistent
calculations), with $U=v=0.14D$. Taking into account Eq.(\ref{eq:Normalized Fano Profile}),
in Fig.\ref{fig:Pic2}, we present the spectral analysis of the bulk
$\text{{LDOS}}=\rho_{0}\frac{(1+q_{J}^{2})}{2}\text{{NFP}}$ as a
function of $\varepsilon$ and the first part of the NFP versus $x$
for several $J$ values. As the dimensionless resonant energy detuning
$x$ is proportional to the deviation from the resonant state $\varepsilon_{d\sigma}=-\frac{U}{2}$
within the valence band, its domain holds for $\varepsilon<0$. Thus,
the second part of Eq.(\ref{eq:Normalized Fano Profile}) as a function
of $\bar{x}$ exhibits a reversed profile, once $\bar{x}=-x$ in the
domain $\varepsilon>0$ (conduction band), where the resonant state
$\varepsilon_{d\sigma}+U=\frac{U}{2}$ resides and $\text{Re}\tilde{\mathcal{G}}_{\sigma}^{0}(\varepsilon>0)=-\text{Re}\tilde{\mathcal{G}}_{\sigma}^{0}(\varepsilon<0)$
fulfills particle-hole symmetry\cite{Pedrosa}. Hence, the dependence
of Eq.(\ref{eq:Normalized Fano Profile}) on $\bar{x}$ is not shown,
for a sake of simplicity.

\begin{figure}[!]
\centering\includegraphics[width=1\columnwidth]{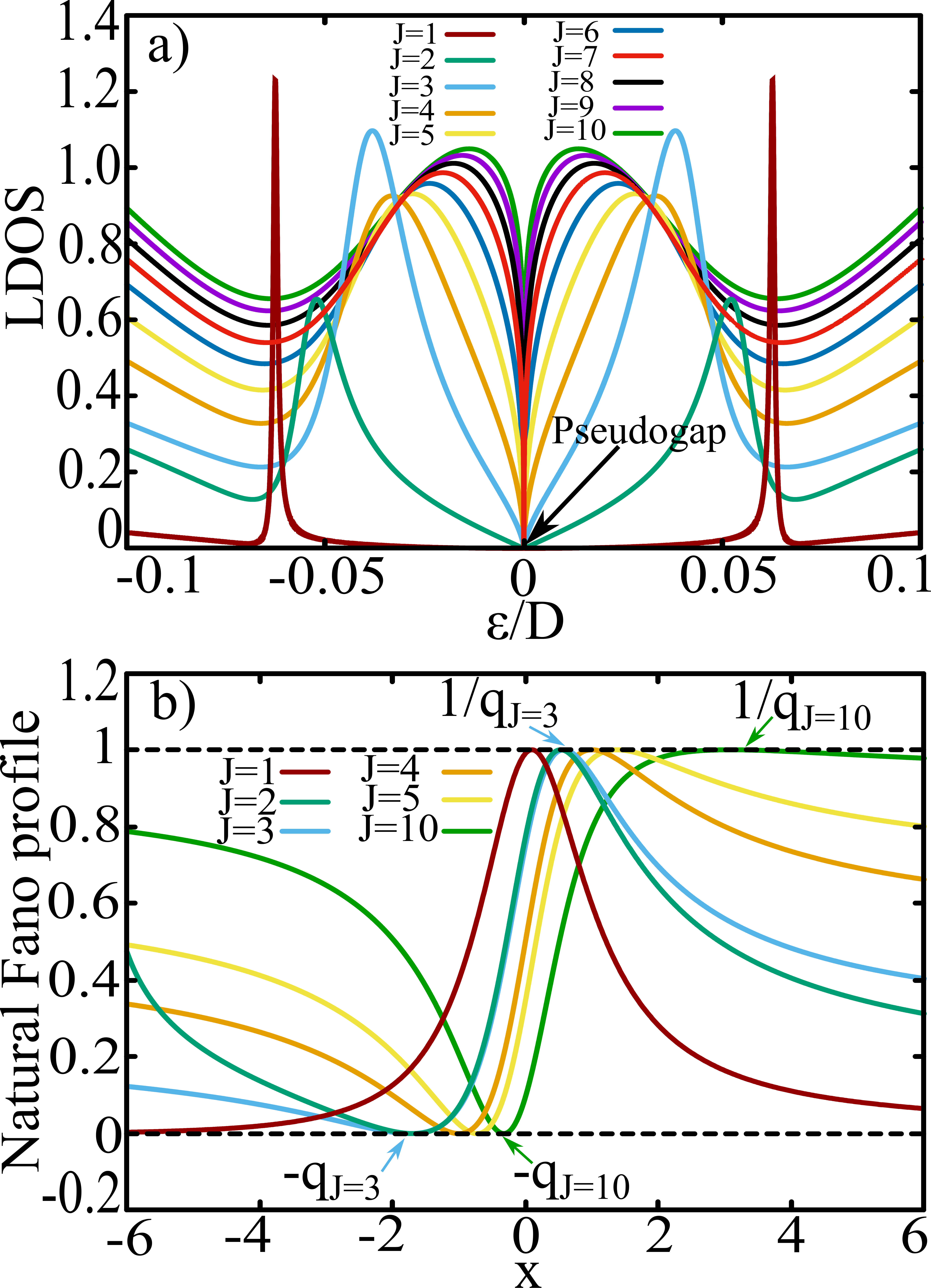}\caption{\label{fig:Pic2} {(Color online) (a) $\text{{LDOS}}=\rho_{0}\frac{(1+q_{J}^{2})}{2}\text{{NFP}}$
{[}Eq.(\ref{eq:Normalized Fano Profile}){]} versus energy $\varepsilon$
in units of the cutoff $D$ for several $J$ values and particle-hole
symmetric regime: $\varepsilon_{d\sigma}=-\frac{U}{2}$ and $w_{x}=w_{\bar{x}}=1/2$
, with $U=v=0.14D$ (see the main text). The increase of $J$ turns
the pseudogap flanked by the impurity resonant states more pronounced
with a sharp dip. (b) Natural Fano profile NFP {[}first part of Eq.(\ref{eq:Normalized Fano Profile}){]}
versus $x$ and dependent on $J$ for $\varepsilon<0.$ In the case
of $\varepsilon>0,$ the profile is just reversed as a function of
$\bar{x}$ and it is not shown, for a sake of simplicity. We clearly
verify that $J$ modulates the Natural Fano profile.}}
\end{figure}

\begin{figure}[!]
\centering\includegraphics[width=1\columnwidth]{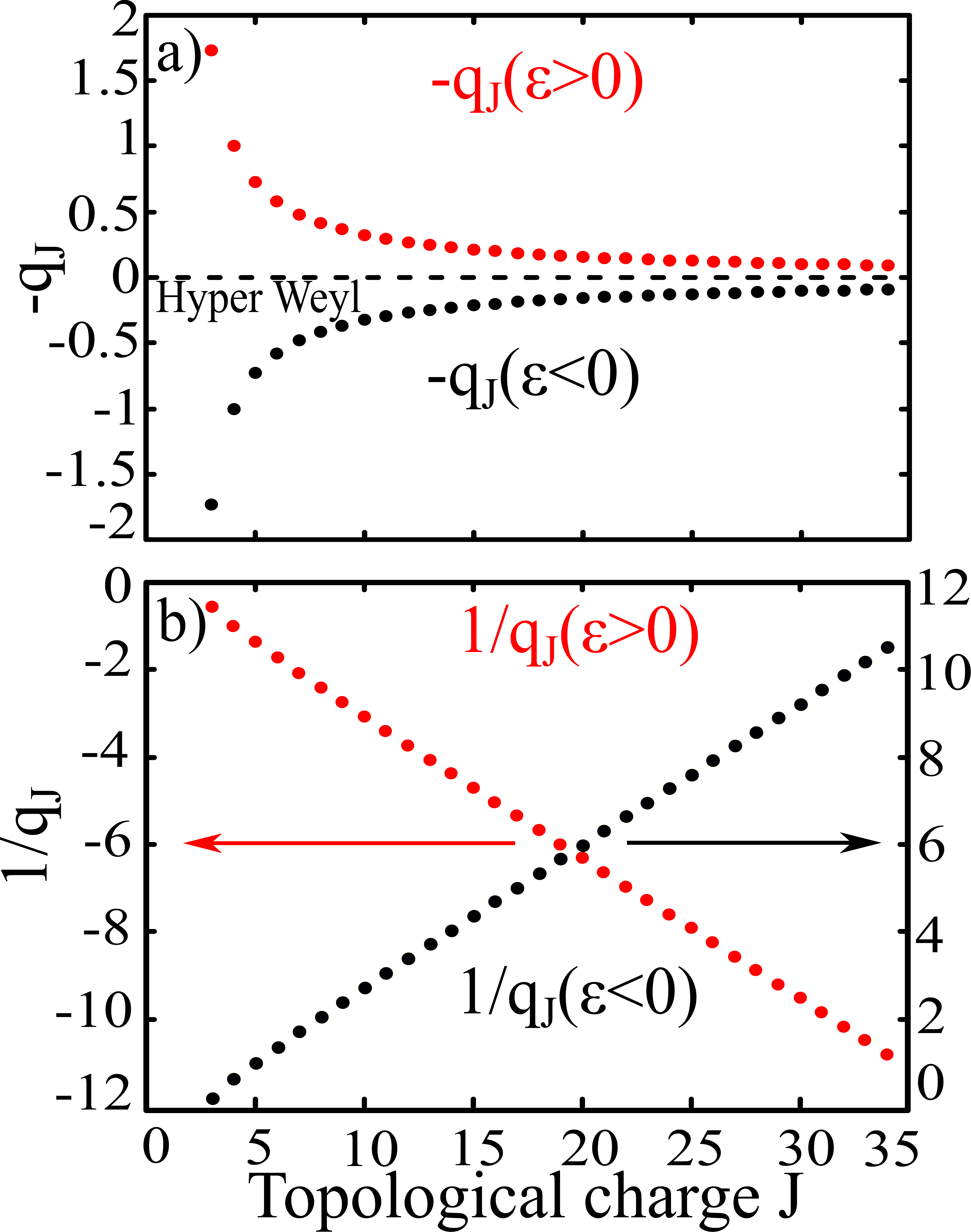}\caption{\label{fig:Pic3} {(Color online) (a) Minimum amplitude position
$-q_{J}$ of the Fano profile from Eq.(\ref{eq:FanoqMaior3}) {[}see
also Figs. \ref{fig:Pic1}(c) and \ref{fig:Pic2}(b){]} with decaying
behavior as a function of $J\geqslant3$ for energies $\varepsilon>0$
and $\varepsilon<0$. (b) The same for the maximum amplitude position
$1/q_{J},$ but with a linear dependence on $J$. Panels (a) and (b)
make explicit that the Fano asymmetry parameter $q_{J}$ given by
Eq.(\ref{eq:FanoqMaior3}) consists of a discretized quantity in the
$J$ term, yielding the topological charge Fano effect.}}
\end{figure}

Counterintuitively, the LDOS in Fig.\ref{fig:Pic2}(a) does not show
Fano profiles around the resonant states as a function of $\varepsilon$
upon changing $J$, as it should occur, despite the Fano parameters
being dictated by Eqs.(\ref{eq:FanoUpTo2}) and (\ref{eq:FanoqMaior3}).
Such a feature arises from the topological charge $J$ and energy
$\varepsilon$ dependencies present in the resonant states broadening
$\Delta=-2v^{2}\text{{Im}\ensuremath{\tilde{\mathcal{G}}_{\sigma}^{0}}}=2\pi v^{2}\rho_{0}$
and quasiparticle dressing term $v^{2}\text{Re}\tilde{\mathcal{G}}_{\sigma}^{0}$
of the impurity. These quantities, in particular, appear in the resonant
energies detuning $x=2(\varepsilon-\varepsilon_{d\sigma}-v^{2}\text{Re}\tilde{\mathcal{G}}_{\sigma}^{0})/\Delta$
and $\bar{x}=2(\varepsilon-\varepsilon_{d\sigma}-U-v^{2}\text{Re}\tilde{\mathcal{G}}_{\sigma}^{0})/\Delta$
entering into Eq.(\ref{eq:Normalized Fano Profile}). From the latter,
we perceive that $x$ and $\bar{x}$ are not linearly proportional
to $\varepsilon$. This characteristic is restored when $\Delta$
and $v^{2}\text{Re}\tilde{\mathcal{G}}_{\sigma}^{0}$ become energy
and $J$ independent, as verified in metallic flat bands near the
Fermi level\cite{Madhavan,Knorr,FanoWithAdatoms3}. Consequently,
solely in this particular situation, the LDOS profile as a function
of $\varepsilon$ shows Fano lineshapes.

Therefore, in the case of multi-Weyl semimetals, one should analyze
$2\text{{LDOS}}/\rho_{0}(1+q_{J}^{2})$, namely, the NFP given by
Eq.(\ref{eq:Normalized Fano Profile}), as a function of the natural
coordinate $x$ or $\bar{x}$ instead of $\varepsilon$, to indeed
perceive the emerging NFP around $x=0$ or $\bar{x}=0$. Such analysis
appears in Fig.\ref{fig:Pic2}(b), where we verify that the increase
of $J$ drives the system from the resonant Fano profile for the case
of single Weyl semimetal $J=1$, towards the hyper Weyl semimetal
with $J\gg1$, which is identified by an antiresonant lineshape. Further,
the Fano minimum and maximum amplitudes positions for $\varepsilon<0$
given by $x=-q_{J}$ and $x=1/q_{J}$, respectively, then appear in
such a figure marked by arrows, just in order to make explicit the
discretized Fano parameter in $J$. However, in Fig.\ref{fig:Pic2}(a)
for the LDOS representation as a function of $\varepsilon$, the role
of $J$ solely lies in the renormalization of the resonant states
towards the Fermi level as $J$ increases, pointing out that the semimetallic
pseudogap becomes characterized by an extremely sharp dip, due to
its spectral power-law $(\varepsilon^{2})^{1/J}$ in Eq.(\ref{eq:ImG0}).

It is worth noting that the discretization observed in Fig.\ref{fig:Pic2}(b)
arises from Eq. (\ref{eq:FanoqMaior3}), which together with Eqs.(\ref{eq:J1}),
(\ref{eq:J2}) and (\ref{eq:FanoUpTo2}) are the most capital ones
of the current work: they encode the topological charge Fano effect
in multi-Weyl systems, once they allow the tuning of the Fano lineshape
by changing the topological charge $J$. Equivalently, according to
the ``bulk-boundary'' correspondence, the pairs of Fermi arcs surface
states present at the system boundaries are fixed by the $J$ value\cite{Dantas},
which also imposes the Fano profile lineshape of the bulk.

With this in mind, we see from Fig.\ref{fig:Pic3}(a) that for $\varepsilon<0$
the limits $-q_{J\rightarrow3}\rightarrow-\tan(C_{2J=6})=-\sqrt{3}$
and $-q_{J\gg1}\rightarrow0$ are reached, while for $\varepsilon>0$
we have $-q_{J\rightarrow3}\rightarrow\tan(C_{2J=6})=\sqrt{3}$ and
$-q_{J\gg1}\rightarrow0$. Interestingly enough for $J=3$, $\left|q_{J=3}\right|=\tan(C_{2J=6})=\sqrt{3}\approx1,732$
and the Fano profile, according to Fig.\ref{fig:Pic2}(b), rises as
asymmetric. Particularly for the $-q_{J}$ decaying behavior with
$J$ reported in Fig.\ref{fig:Pic3}(a), we highlight that such a
feature is connected straightforwardly to the system band-structure.
As stated previously, the band-structure saturates into one characteristic
for the hyper Weyl semimetal-type with $J\gg1$, as depicted in Fig.\ref{fig:Pic1}(d).
As an aftermath, if the shape generated by the dispersion $\varepsilon_{\textbf{k}s}^{\pm}$
from Eq.(\ref{eq:DispersionRelation}) remains unchanged by increasing
$J\gg1$ {[}Fig.\ref{fig:Pic1}(d){]}, so does the Fano parameter,
which attains to $\left|q_{J\gg1}\right|\rightarrow0$, once it depends
on $\varepsilon_{\textbf{k}s}^{+}$ via the GF $\tilde{\mathcal{G}}_{\sigma}^{0}$
of the pristine host. As a result, the antiresonant Fano profile becomes
the hallmark of hyper Weyl semimetals.

In Fig.\ref{fig:Pic3}(b), we show the corresponding behavior for
$1/q_{J}$, which is linear instead. Note that $1/q_{J}$ for $\varepsilon<0$
follows an increasing linear trend, while for $\varepsilon>0$ it
is the opposite. This reflects the own particle-hole symmetry characteristic
of the Fano parameter and it occurs because the multi-Weyl points
and Fermi level are energy-degenerate. Thereby, the band-structure
also has particle-hole symmetry, as well as the $1/q_{J}$ quantity.
Most importantly, both the $-q_{J}$ and $1/q_{J}$ behaviors as functions
of $J$ make explicit that the Fano parameter is discretized, thus
characterizing the topological charge Fano effect in multi-Weyl semimetals.

\subsection{Experimental proposal to detect the topological charge Fano effect}

\begin{figure}[!]
\centering\includegraphics[width=1\columnwidth]{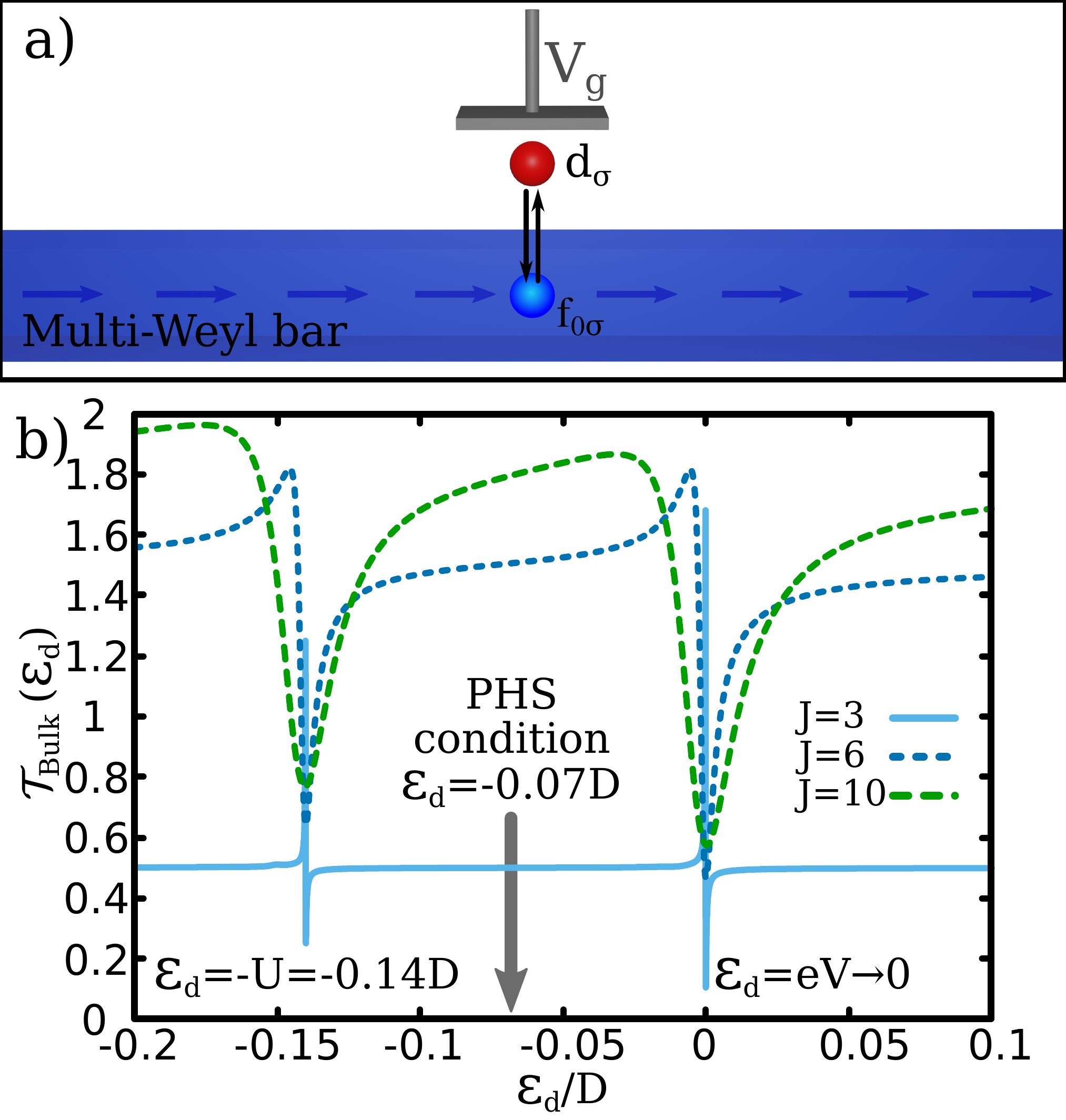}\caption{\label{fig:Pic4} {{(Color online) (a) Sketch of
the multi-Weyl bar device, where an impurity appears side-coupled.
The arrows denote the current direction from the source to drain leads
(not depicted). This system consists of an experimental proposal for
detecting the topological charge Fano effect. (b) The bulk transmittance
$\mathcal{T}_{\text{{Bulk}}}(\varepsilon_{d})=\text{{NFP}}$ versus
$\varepsilon_{d}$ for several $J$ values is determined via Eq.(\ref{eq:Normalized Fano Profile}),
with $\varepsilon=\text{{eV}}\rightarrow0$ as a small bias-voltage.
The $\varepsilon_{d}$ degree is tunable by a gate voltage $V_{g}$
attached to the impurity, thus leading to a transmittance with Fano
profiles nearby the energies $\varepsilon_{d}=-U$ and $\varepsilon_{d}=\text{{eV}.}$
Around them, we can apply Eq.(\ref{eq:Normalized Fano Profile}) to
experimental data, extract the Fano asymmetry parameter $q_{J}$ and
estimate via Eq.(\ref{eq:FanoqMaior3}) the topological charge $J.$
The middle point between such energy positions corresponds to the
particle-hole symmetry (PHS), wherein the condition $\varepsilon_{d}=-\frac{U}{2}$
is fulfilled.}}}
\end{figure}

{From the spectral analysis performed so far, we perceive
that the NFP of Eq.(\ref{eq:Normalized Fano Profile}) requires natural
coordinates to be viewed, such as $x$ and $\bar{x}$ from Eqs.(\ref{eq:naturalx})
and (\ref{eq:naturalxbar}), respectively. This characteristic relies
in the fact that both the latter expressions are highly non-linear
functions in the $\varepsilon$ degree, which turns the detection
of the Fano profiles encoded by Eq.(\ref{eq:Normalized Fano Profile})
a hard challenge by varying $\varepsilon$. However, we propose an
alternative path to overcome such an experimental obstacle. We begin
with by noticing that if we consider the impurity level $\varepsilon_{d\sigma}=\varepsilon_{d}$
in Eq.(\ref{eq:Imp}) a tunable parameter, while $\varepsilon,$ $\text{Re}\tilde{\mathcal{G}}_{\sigma}^{0}$
and $\text{Im}\tilde{\mathcal{G}}_{\sigma}^{0}$ as constant numbers,
we finally gain the desired linear dependence of $x$ and $\bar{x},$
not with $\varepsilon,$ but with $\varepsilon_{d}$ instead. Thus,
Eq.(\ref{eq:Normalized Fano Profile}) as a function of $\varepsilon_{d}$
is expected to show Fano profiles, once $\varepsilon_{d}$ rises as
the natural coordinate for the emanation of the Fano profile. The
tunability of $\varepsilon_{d}$ to become feasible from an experimental
perspective needs a remake of Fig.\ref{fig:Pic1}(a) into the transport-type
device of Fig.\ref{fig:Pic4}(a), which we introduce as the multi-Weyl
bar. }

{The multi-Weyl bar consists of a quasilinear bulk
system with an external impurity, where this side-coupled impurity
is supposed to overlap with both the surface and bulk states of the
system. A similar approach was done to the electron channel treated
in the quantum wire theoretically explored in Ref.\cite{SideCoupledQD}.
In such a work, one of us derived a transmittance formula of an impurity
side-coupled to the electron channel and found Fano profiles in the
zero-bias conductance by tuning $\varepsilon_{d},$ due to the assumption
of a gate voltage $V_{g}$ attached to this impurity. It was found
that the conductance reflects the system bulk properties via the transmittance
coefficient, hereby with the shorthand notation $\mathcal{T}_{\text{{Bulk}}}(\varepsilon_{d})=\frac{\mathcal{G}}{\mathcal{G}_{0}}$,
where $\mathcal{G}$ is the zero-bias conductance at $T\ll T_{\text{{K}}}\rightarrow0$
and $\mathcal{G}_{0}=\frac{e^{2}}{h}$ stands for the conductance
quantum. The bulk properties come up, once $\mathcal{T}_{\text{{Bulk}}}(\varepsilon_{d})$
depends upon the spectral density of the host site described by the
fermionic operator connected to the impurity. Thus, in our system,
this corresponds to Eq.(\ref{eq:fo}) for $f_{0\sigma}$ and consequently,
it allows, together with Eq.(\ref{eq:Normalized Fano Profile}) for
the NFP and the quantum transport formalism developed in Ref.\cite{SideCoupledQD},
to derive the equality $\mathcal{T}_{\text{{Bulk}}}(\varepsilon_{d})=\text{{NFP}}.$
As the NFP is bounded by $2$, this upper limit represents the maximum
of the transmittance by accounting for the two spin channels.}

{However, we should adapt carefully the transport
formalism done in Ref.\cite{SideCoupledQD} for a quantum wire to
multi-Weyl semimetals in the geometry of the bar depicted in Fig.\ref{fig:Pic4}(a).
First, multi-Weyl semimetals have pseudogap {[}Eq.(\ref{eq:ImG0}){]}
and it does not make sense to perform a zero-bias analysis, but this
can be easily solved by placing $\mathcal{T}_{\text{{Bulk}}}(\varepsilon_{d})$
slightly off the charge neutrality point $\varepsilon=0$, namely
$\varepsilon=\text{{eV}}\rightarrow0,$ being $\text{{eV}}$ a small
bias-voltage in which the system conducts. This fixes $\varepsilon,$
$\text{Re}\tilde{\mathcal{G}}_{\sigma}^{0}$ and $\text{Im}\tilde{\mathcal{G}}_{\sigma}^{0}$
at $eV\rightarrow0$ in Eqs.(\ref{eq:naturalx}) and (\ref{eq:naturalxbar})
for $x$ and $\bar{x}$, respectively, thus restoring the highly desired
linearity of these quantities with $\varepsilon_{d},$ which is necessary
for the Fano profiles to appear. Second, the practical realization
of the multi-Weyl bar implies in a finite system, where the Fermi
arcs surface states contribute inevitably to the total transmittance,
together with the bulk states with $\text{{eV}}\rightarrow0.$ Despite
this present characteristic of the experimental proposal, as Fermi
arcs surface states are topologically protected\cite{Hasan2021,Hasan2017,Yan,Zheng,Armitage,Hu,Marques,MizobataWeyl},
in opposite to the bulk states, the Fano patterns in the total transmittance
as a function of $\varepsilon_{d}$ are expected to have the latter
as their source. This means that topologically protected states are
supposed to stay robust under external perturbations, in particular,
those that do not break the symmetry that protect such states. In
this manner, these states become immune to Fano interference. Here,
by changing $\varepsilon_{d}$, the particle-hole symmetry (PHS) of
the Hamiltonian of Eq.(\ref{eq:TotalH}) breaks down, but it does
not unprotec topologically the Fermi arcs surface states, once they
are not protected by such a symmetry. Consequently, the Fano patterns
in the total transmittance are expected to be dictated by $\mathcal{T}_{\text{{Bulk}}}(\varepsilon_{d}),$
which for $\varepsilon=\text{{eV}}\rightarrow0,$ leads to Fano profiles
around $\varepsilon_{d}=-U$ and $\varepsilon_{d}=\text{{eV}},$ as
shown in Fig.\ref{fig:Pic4}(b) for several $J$ values. }

{We stress that the quantification concerning whether
the total transmittance depends weakly on the Fermi arcs surface states
is not the focus of the main analysis of the current work. Thus, further
investigation into the dependence degree of the Fano interference
on the Fermi arcs surface states will be addressed elsewhere. It is
worth emphasizing that the proposed device depicted in Fig.\ref{fig:Pic4}(a)
points out a way to induce Fano lineshapes in the total transmittance,
just by changing $\varepsilon_{d}$ for fixed small bias-voltage $\text{{eV}}\rightarrow0.$
Experimentally speaking, from one detected Fano lineshape, we can
extract the Fano asymmetry parameter $q_{J}$ via Eq.(\ref{eq:Normalized Fano Profile})
and by employing Eq.(\ref{eq:FanoqMaior3}), for instance, determine
the topological charge $J$. }

{To summarize, for $\mathcal{T}_{\text{{Bulk}}}(\varepsilon_{d})$
versus $\varepsilon_{d},$ the Fano lineshape in Fig.\ref{fig:Pic4}(b)
is also modulated by $J$ and obeys the same trend previously observed
in Fig.\ref{fig:Pic2}(b). Equivalently, the deeper meaning of the
``bulk-boundary'' correspondence applied to the here proposed Fano
effect is the following: the greater the amount of Fermi arcs surface
states given by $J$ at the system boundaries, the more the Fano profile
will be antiresonant within the bulk.}

\section{Conclusions}

In this work, we determine the Fano asymmetry parameter for a single
impurity coupled to a multi-Weyl semimetal and introduce the concept
of topological charge Fano effect. According to the ``bulk-boundary''
correspondence, which states that the number of Fermi arcs at the
boundaries of a finite size system is determined by the magnitude
of the topological charge, known from its bulk version with infinite
size, we then reveal the modulation of the system Fano profile, due
to the bulk LDOS, by such surface states. This can be emulated in
our theoretical framework by the tuning of the topological charge
value, which allows the Fano profile to change from the resonant pattern
for single Weyl semimetal, towards the antiresonant Fano lineshape,
which identifies hyper Weyl semimetals. Additionally, for the maximum
allowed protected case by the rotational symmetry group $C_{2J=6}$,
namely, the triple Weyl semimetal $J=3$ and rotational angle defined
by $C_{2J}\equiv(360^{\circ}/2J)$, we predict the absolute Fano parameter
$\left|q_{J=3}\right|=\tan(C_{2J=6})$ and an asymmetric Fano profile.
{Additionally, we indicate a quantum transport setup
where we expect that the here proposed Fano effect could be present.}

\section{Acknowledgments}

We thank the Brazilian funding agencies CNPq (Grants. Nr. 302887/2020-2,
308410/2018-1, 311980/2021-0, 305738/2018-6, 311366/2021-0, 305668/2018-8 and 308695/2021-6),
Coordenação de Aperfeiçoamento de Pessoal de Nível Superior - Brasil
(CAPES) -- Finance Code 001, the São Paulo Research Foundation (FAPESP;
Grant No. 2018/09413-0) and FAPERJ process Nr. 210 355/2018. LSR and
IAS acknowledge support from the Icelandic Research Fund (project
``Hybrid polaritonics''). IAS also acknowledges support from the
Program Priority 2030. LSR thanks ACS and Unesp for their hospitality.}

\end{document}